\documentclass[article]{JHEP3} 

\usepackage{epsfig,multicol,amsmath}

\newcommand\fverb{\setbox\pippobox=\hbox\bgroup\verb}
\newcommand\fverbdo{\egroup\medskip\noindent%
			\fbox{\unhbox\pippobox}\ }
\newcommand\fverbit{\egroup\item[\fbox{\unhbox\pippobox}]}
\newbox\pippobox
\def\d2bar{$\overline{\mbox D2}$}
\title{Lineal Trails of D{\bf 2}-$\overline{\mbox D\bf{2}}$ Superstrings}
\author{Jin-Ho Cho, Phillial Oh, Cheonsoo Park, and Jonghyeon Shin\\
	BK21 Physics Research Division and Institute of Basic Science\\
 Sungkyunkwan University, Suwon 440-746, Korea\\
	E-mail: \email{jhcho@taegeug.skku.ac.kr}, \email{ploh@dirac.skku.ac.kr}, \email{cspark@newton.skku.ac.kr}, \email{gempyonn@korea.com}}

\preprint{\hepth{0312094}}	

\abstract{We study the superstrings suspended between a D2- and a \d2bar-brane. We quantize the string in the presence of some general configuration of gauge fields over the (anti-)D-brane world volumes. The interstring can move only in a specific direction that is normal to the difference of the electric fields of each (anti-)D-branes. Especially when the electric fields are the same, the interstring cannot move. We obtain the condition for the tachyons to disappear from the spectrum.}


\begin{document} 

\section{Introduction}
Contrary to our naive conceptions, there are some configurations of D-branes for which supersymmetry is enhanced by turning on noncommutativity. Especially  for odd values of $k$, D0-D(2$k$)-brane system can be supersymmetric in the presence of suitable background Born-Infeld (BI) magnetic fields \cite{park,witten}. The BI electric fields add more possibilities of the supersymmetries to the D-brane composite systems \cite{cho-oh}. In the T-dual description, they can be connected to two D-strings at angle (scissors) and in a relative motion. The 8 supersymmetries are preserved when the intersecting point of the scissors is moving at the speed of light, thus called null scissors \cite{bachas} \cite{myers} \cite{chen}.  

Supertube provides another nontrivial example that is stabilized by the interplay of the electric field and the magnetic field \cite{townsend}. Its world volume theory is a noncommutative gauge theory in the rotating frame \cite{lee}. The configuration is T-dual to the super D-helix \cite{cho-oh2}, which is a nontrivial coiled supersymmetric D-string in motion with the speed of light. Furthermore, a higher dimensional D-brane with a plane electromagnetic wave on its world volume is supersymmetric \cite{bachas2}.

Even an extreme model of a D2-\d2bar pair can be stabilized by a keen adjust of the critical valued electric field ($2\pi\alpha'E=1$) and the magnetic fields over the (anti-)brane world volumes \cite{bak1}. In the string theory context, this can be understood as the removal of tachyonic interstrings between the D2- and \d2bar-brane by the specific background gauge fields acting on two ends of those strings \cite{bak}. An extension to a D3-$\overline{\mbox D3}$ pair is also possible \cite{jabbari}.

The purpose of this paper is to study the generality of the critical value of the electric field in these kind of stabilizing mechanism. This was explored in \cite{cho-oh}, in different context, by counting the number of the spacetime supercharges preserved by various D-$\overline{\mbox{D}}$ configurations. The critical value of the electric field is so crucial to the supersymmetries that only the null scissors and its T-dual configuration were the exceptional supersymmetric models involving noncritical values of the electric field. 

In this paper we study a superstring suspended between a D2-brane and a \d2bar-brane with an arbitrary distribution of constant electric fields and constant magnetic fields on their world-volumes. We find the condition for the tachyons to disappear from the string spectrum. The condition is 
\begin{eqnarray}\label{notachyon}
\left(1-\vec{e}\cdot\vec{e}+\vec{b}\cdot\vec{b}\right)\left(1-\vec{e}'\cdot\vec{e}'+\vec{b}'\cdot\vec{b}'\right)-\left(1-\vec{e}\cdot\vec{e}'+\vec{b}\cdot\vec{b}'\right)^2\rightarrow0,
\end{eqnarray}
where $(\vec{e},\,\vec{b})$ and $(\vec{e}',\,\vec{b}')$ are sets of the electric and the magnetic fields over each (anti-)D2-brane. This condition is satisfied by all known examples of (anti-)D2-brane composites including the null scissors. As a byproduct, we show that the interstrings can move in parallel with same speed. The momentum measured in the open string metric agrees with the notion of the energy of the spectrum. 

One can find many seminal papers on the open strings coupled to the background gauge fields, even prior to D-branes \cite{tseytlin, callan1,callan2}. In the modern langauge, they mostly ellaborate on the open strings on a single sort of D-branes. We should also mention the paper by Bachas and Porrati, who studied the pair creation of open strings in an electric field in the context of the unoriented string theories \cite{bachas3}.
 
The organization of the paper is as follows. In the next section, we obtain the solution for an interstring supended over a D2-\d2bar pair in the presence of a general configuration of the gauge fields. The solution allows fractional modes besides the integer modes one encounters in the ordinary solutions of strings. Those modes in the directions concerned with the gauge fields become intermingled and are not independent in generic cases. In Sec. III, we quantize the system by the symplectic method developed in Ref. \cite{chu1} and \cite{chu2}. Especially we work out all the constraints prior to taking the inverse of the symplectic two form, which makes the cumbersome constraint analysis unnecessary. As the result, we show in Sec. IV, the space-time noncommutativity as well as the space-space noncommutativity. Sec. V shows the way to avoid the tachyons in the string spectrum. We show that the appearance of the novel fractional modes compensate for the absence of the spatial longitudinal integer modes. The massless R-spectrum matches exactly with the NS-spectrum after GSO projection only when the condition (\ref{notachyon}) is satisfied. In Sec. VI, we show that all the interstrings follow parallel lineal trails with the same momentum. Sec. V closes the paper with some remarks on the consistency with other results and with some outlooks for future works.

\section{Setup}

Let us consider an interstring over a D2-\d2bar pair. Each end point is interacting with the background gauge fields on each brane world-volumes. Denoting the string world-sheet coordinates of these ends as $\sigma=0$ and $\sigma=\pi$ respectively, we give the background gauge fields $B^{(\sigma)}_{\mu\nu}$ as
\begin{eqnarray}
(B^{(0)}_{\mu\nu})=\left(
\begin{matrix}
0  & 0 &  e \cr
0  & 0 & -b \cr
-e & b & 0
\end{matrix}
\right),\qquad
(B^{(\pi)}_{\mu\nu})=
\left(\begin{matrix}
0 & e'\sin\theta & e'\cos\theta \cr
-e' \sin\theta & 0 & b' \cr
-e' \cos\theta & -b'& 0
\end{matrix}\right).
\end{eqnarray}
Note that $B^{(\sigma)}_{\mu\nu}$ is dimensionless and is related to the BI field strength as $B=2\pi\alpha'F$.

\begin{figure}
\epsfbox{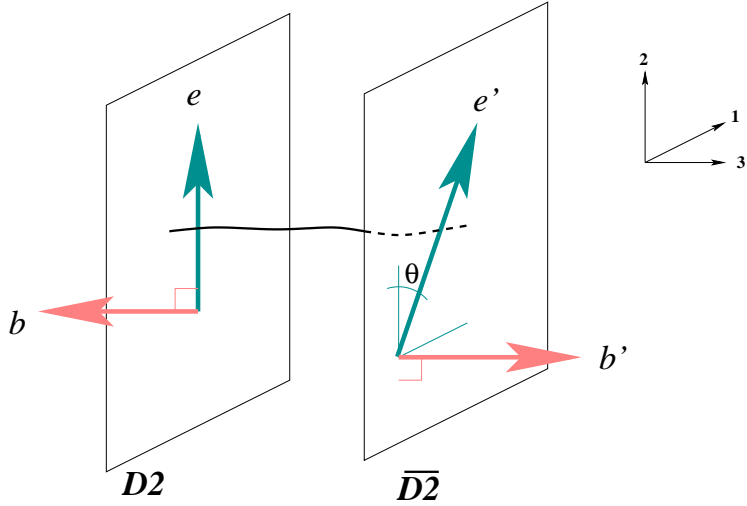}
\caption{A superstring is suspended between a D2-brane and a \d2bar-brane posed in parallel. We assume different values of electric fields $\vec{e}$ and $\vec{e}'$ on each world volumes of D-branes. The magnetic fields $\vec{b}$ and $\vec{b}'$ are normal to the world-volumes but their magnitudes are different.}
\end{figure}

In the flat spacetime, the Neveu-Schwarz superstring is described by the action;
\begin{eqnarray}\label{action}
S&=&\left.-\frac{1}{4\pi\alpha'} \int d^2\sigma\right[\partial_\alpha X^\mu\partial^\alpha X^\nu\eta_{\mu\nu}+\bar{\psi}^\mu\gamma^\alpha\partial_\alpha\psi^\nu\eta_{\mu\nu}\nonumber\\
&&\left.\quad+\left(2 A^{(\sigma)}_\mu\dot{X}^\mu+\frac{1}{2}\bar{\psi}^\mu\gamma^1\psi^\nu B^{(\sigma)}_{\mu\nu}\right)\Bigl(\hat{\delta}(\sigma)-\hat{\delta}(\sigma-\pi)\Bigr)\right],
\end{eqnarray} 
where $\gamma^0=i\sigma^2$, $\gamma^1=\sigma^1$. The delta function with hat is defined on the compact space $[0,\,\pi]$. The second line contains the boundary terms which determine the boundary conditions as follow. The longitudinal directions are given by the mixed boundary conditions,
\begin{eqnarray}\label{bdboson}
\left.\left(E^{(\sigma)}_{\nu\mu}\partial_+X^\nu-E^{(\sigma)}_{\mu\nu}\partial_-X^\nu\right) \right|_{\sigma =0, \pi}
  = 0 \qquad (\mu , \nu = 0, 1, 2).
\end{eqnarray}
Here we used the abbreviation $E^{(\sigma)}_{\mu\nu}=\eta_{\mu\nu} +B^{(\sigma)}_{\mu\nu}$, and the lightcone coordinates, $\sigma^+\equiv\tau+\sigma$ and $ \sigma^-\equiv\tau-\sigma$. The other directions are governed by Dirichlet conditions $X^{\mu} |_{\sigma= 0,\pi} =0$, but for the direction $X^3$, along which  the D2- and \d2bar-branes are separated by a distance $l$, thus $X^3 |_{\sigma =0} = 0$ and $X^3 |_{\sigma =\pi} =l$.  

As for the fermions, there are two consistent boundary conditions, corresponding to Ramond- (R-) and Neveu-Schwarz- (NS-)sector respectively:
\begin{eqnarray}\label{bdfermion}
&&\left\{
\begin{array}{l}
E^{(\sigma)}_{\nu\mu}\psi^{\nu}_{+} |_{\sigma =0}
  = E^{(\sigma)}_{\mu\nu}\psi^{\nu}_{-} |_{\sigma =0}, \\
E^{(\sigma)}_{\nu\mu}\psi^{\nu}_{+} |_{\sigma =\pi}
  =\pm E^{(\sigma)}_{\mu\nu}\psi^{\nu}_{-} |_{\sigma =\pi},
\end{array}
\quad(\mu , \nu = 0, 1, 2),\right.\nonumber\\
&&
\left\{
\begin{array}{l}
\psi^{\mu}_{+} |_{\sigma =0} =-\psi^{\mu}_{-} |_{\sigma =0}, \\
\psi^{\mu}_{+} |_{\sigma =\pi} =\mp\psi^{\mu}_{-} |_{\sigma =\pi},
\end{array}\qquad\qquad\,\,\,\,
(\mu = 3,\cdots ,9).\right.
\end{eqnarray}
The upper/lower sign is for the R-sector/NS-sector. The Majorana spinor $\psi^\mu=(\psi^\mu_+,\,\psi^\mu_-)$ has been rewritten in components.

The solutions are nontrivial for the longitudinal directions. Let us first consider the bosonic part. Imposing the boundary conditions at $\sigma=0$, one can solve the equations of motion and get
\begin{eqnarray}\label{solboson}
X_{\mu}&=&\frac{x_{\mu}}{2}+E^{(0)}_{\mu\nu}\left[a^\nu_0\sigma^++\sum\limits_{n\ne 0}\frac{i}{n}a^\nu_ne^{-in\sigma^+}+\sum\limits_{n}\frac{i}{2(n+\kappa)}\left(a^\nu_{n+\kappa}e^{-i(n+\kappa)\sigma^+}-\bar{a}^\nu_{-n-\kappa}e^{i(n+\kappa)\sigma^+}\right)\right]\nonumber\\
&+&\frac{x_{\mu}}{2}+E^{(0)}_{\nu\mu}\left[a^\nu_0\sigma^-+\sum\limits_{n\ne 0}\frac{i}{n}a^\nu_ne^{-in\sigma^-}+\sum\limits_{n}\frac{i}{2(n+\kappa)}\left(a^\nu_{n+\kappa}e^{-i(n+\kappa)\sigma^-}-\bar{a}^\nu_{-n-\kappa}e^{i(n+\kappa)\sigma^-}\right)\right].\nonumber\\
\end{eqnarray}
Here we allowed some generality on mode expansion. In the fractional modes, $0<\kappa<1$. This is necessary in order to make nontrivial solutions satisfying the boundary conditions at the other end. We introduced the complex conjugate parts $\bar{a}^\nu_{-n-\kappa}=(a^\nu_{n+\kappa})^*$ to make $X_\mu$ real. Likewise $a^\nu_{-n}=(a^\nu_n)^*$.

One might worry about these excessive oscillator modes, $a^\mu_{n+\kappa}$ and $\bar{a}^\mu_{-n-\kappa}$ along the longitudinal directions $(\mu=0, 1, 2)$, which are absent in the conventional strings in the trivial background. In the end (Sec. 5), we will see that only one set of oscillator modes for each longitudinal directions are physical.

In order for the integer modes to satisfy the boundary condition at $\sigma=\pi$, 
\begin{eqnarray}\label{rel1}
\left(\begin{matrix}\label{integer}
0 & -e'\sin{\theta} & e-e'\cos{\theta}\cr
e'\sin{\theta} & 0 & -b-b'\cr
e'\cos{\theta}-e & b+b' & 0 
\end{matrix}\right)
\left(\begin{matrix} 
a^0_n\cr a^1_n\cr a^2_n
\end{matrix}\right)=0,\qquad (n=0, \pm1, \pm2,\cdots ).
\end{eqnarray}
Therefore the components are unconstrained when $\vec{e}=\vec{e}'$ and $\vec{b}=\vec{b}'$.\footnote{For the notational simplicity, we use the 3-dimensional vector notation for the electric fields and the magnetic fields: $\vec{e}=(B^{(0)}_{01}, B^{(0)}_{02}, 0)$, $\vec{b}=(0, 0, B^{(0)}_{12})$, $\vec{e}'=(B^{(\pi)}_{01}, B^{(\pi)}_{02}, 0)$, $\vec{b}'=(0, 0, B^{(\pi)}_{12})$.} However in generic cases, the above condition tells us that the components, $a^0_n,\,a^1_n,\,a^2_n$ depend on one another.
One can classify the cases as follows:
i) $\vec{e}=\vec{e}',\quad \vec{b}\ne\vec{b}'$ : $a^0_n$ is unconstrained but $a^1_n=a^2_n=0$.
ii) $\vec{e}\ne\vec{e}',\quad \vec{b}\ne\vec{b}'$ : $a^0_n,\,a^1_n,\,a^2_n$ are linearly depend on one another.
iii) $\vec{e}=\vec{e}',\quad \vec{b}=\vec{b}'$ : $a^0_n,\,a^1_n,\,a^2_n$ are independent.
iv) $\vec{e}\ne\vec{e}',\quad \vec{b}=\vec{b}'$ : $a^0_n=0$ and the vector $(a^1_n,\,a^2_n)$ is normal to the vector $\vec{e}-\vec{e}'$.
From hereon we will specialize to the first two cases, i) and ii) by assuming $\vec{b}\ne \vec{b}'$. The other cases iii) and iv) need separate careful analyses, which will be covered elsewhere \cite{future}.

The boundary condition at $\sigma=\pi$ regulates the fractional modes too:
\begin{eqnarray}\label{kappa}
\left[i\tan{\kappa\pi}\left(\eta_{\mu\lambda}-B^{(\pi)}_{\mu\rho}\eta^{\rho\nu}B^{(0)}_{\nu\lambda}\right)-\left(B^{(0)}_{\mu\lambda}-B^{(\pi)}_{\mu\lambda}\right)\right]a^\lambda_{n+\kappa}=0.
\end{eqnarray} 
In order for this equation to have nontrivial solutions for $a^\lambda_{n+\kappa}$, the determinant of the prefactor should vanish. This determines $\beta\equiv \tan{\kappa\pi}$ to be
\begin{eqnarray}\label{kappa}
\beta=0,\quad\mbox{or}\quad \beta^2=\frac{\left(1-\vec{e}\cdot\vec{e}+\vec{b}\cdot\vec{b}\right)\left(1-\vec{e}'\cdot\vec{e}'+\vec{b}'\cdot\vec{b}'\right)-\left(1-\vec{e}\cdot\vec{e}'+\vec{b}\cdot\vec{b}'\right)^2}{\left(1-\vec{e}\cdot\vec{e}'+\vec{b}\cdot\vec{b}'\right)^2}.
\end{eqnarray} 
Especially when $\beta=0$, the fractional modes satisfy the same equation as (\ref{integer}), thus their components are mutually dependent in the same way as the integer modes. In more general situation, we may have nontrivial solutions for $\beta$ if $\left(1-\vec{e}\cdot\vec{e}+\vec{b}\cdot\vec{b}\right)\left(1-\vec{e}'\cdot\vec{e}'+\vec{b}'\cdot\vec{b}'\right)$ is larger than $\left(1-\vec{e}\cdot\vec{e}'+\vec{b}\cdot\vec{b}'\right)^2$. In this latter case, the components of $a^\mu_{n+\kappa}$ are related with one other in somewhat complicated form.
\begin{eqnarray}\label{rel2}
a^{1}_{n+\kappa}=\frac{r+is}{p+iq}~a^{0}_{n+\kappa},\qquad a^{2}_{n+\kappa}=\frac{t+iu}{p+iq}~a^{0}_{n+\kappa},
\end{eqnarray}
where
\begin{eqnarray}
p&=&\left(1-bb'\right)\left(1-ee'\cos\theta-bb'\right)\beta^2-\left(b+b'\right)^2,\nonumber\\
q&=&\left(b+b'\right)ee'\sin\theta\,\beta,\nonumber\\
r&=&-eb'\left(1-ee'\cos\theta-bb'\right)\beta^2+\left(b+b'\right)\left(e'\cos\theta-e\right),\nonumber\\
s&=&\left[b\left(b+b'\right)-\left(1-e^2+b^2\right)\right]e'\sin\theta\,\beta,\nonumber\\
t&=&-(b+b')e'\sin\theta,\nonumber\\
u&=&\left[e\left(b'^2+1\right)-e'\cos\theta\left(1-bb'\right)\right]\beta.
\end{eqnarray}
The explicit expression is not important in our later discussion. The only thing we have to note here is that $q,\,s,\,u$ concerned with the imaginary parts disappear when $\beta=0$. Therefore the relations among $a^\mu_{n+\kappa}, (\mu=0,\,1,\,2)$ become the same as those of integer part, as they ought to be. 

Let us move on to the fermionic part. The solutions $\psi^\mu_{\pm}$, satisfying the boundary condition at $\sigma=0$, look very similar to the expressions of $\partial_{\pm}X^\mu$. The only difference is that the mode frequency $n$ is replaced by $r$, that is integer valued in the R-sector and half integer valued in the NS-sector. 
\begin{eqnarray}\label{solfermion}
\psi_{-\mu}&=&\sum_r E^{(0)}_{\lambda\mu}
\left[h^\lambda_r e^{-ir\sigma^-}+\frac{1}{2}(h^\lambda_{r+\kappa} e^{-i(r+\kappa)\sigma^-}+\bar{h}^\lambda_{-r-\kappa}~ e^{i(r+\kappa)\sigma^-})\right]\nonumber\\
\psi_{+\mu}&=&\sum_r E^{(0)}_{\mu\lambda}
\left[h^\lambda_r e^{-ir\sigma^+}+\frac{1}{2}(h^\lambda_{r+\kappa} e^{-i(r+\kappa)\sigma^+}+\bar{h}^\lambda_{-r-\kappa}~ e^{i(r+\kappa)\sigma^+})\right],
\end{eqnarray}
where $\bar{h}^\lambda_{-r-\kappa}~=(h^\lambda_{r+\kappa})^*$ and $r\in Z\!\!\!\!Z$ for the R-sector while $r\in Z\!\!\!\!Z+1/2$ for the NS-sector. One might naively think that there is an interesting merge of the NS-sector and the R-sector when $\kappa\rightarrow 1/2$. However, no such a thing happens. We will discuss about this later.
As with bosonic case, the fermionic part undergoes nontrivial relations among mode coefficients upon the imposition of the boundary conditions on the other end of the string $(\sigma=\pi)$. The precise form of those dependencies turns out to be the same as the form shown in Eq. (\ref{rel1}) and in Eq. (\ref{rel2}), but with the replacements of $a$'s with $h$'s. 

\section{Quantizing the Interstrings in a General Background of Gauge Fields}

In this section we quantize the interstring discussed above and study its motion. It is not trivial to quantize the system because the mixed boundary conditions corresponds to the constraints about the coordinate fields, $X^\mu$ and their  conjugate momenta $\Pi_\mu$ at two string ends. It looks highly nontrivial to quantize the system that has constraints just at two points. The idea to solve this complexity is to go over to the momentum space. See Ref. \cite{chu1, chu2} for details. Inserting the solutions (\ref{solboson}) and (\ref{solfermion}) into the action (\ref{action}), we would get a system composed of infinite point particles upon the integration over $\sigma$. The boundary conditions can be incorporated into the system through the relations (\ref{rel1}) and (\ref{rel2}) and their fermion correspondents. However cumbersome it is, in principle, one can quantize the system. 

The Poisson structure is can be read off from the symplectic two-form,
\begin{eqnarray}
\Omega=\int^\pi_0d\sigma~~\delta{\Pi_X}_\mu\wedge\delta X^\mu-\int^\pi_0d\sigma~~\delta{\Pi_\psi}_\mu\wedge\delta\psi^\mu,\nonumber
\end{eqnarray}
where we denoted the exterior derivative in the phase manifold as `$\delta$' in order to avoid the confusion with that on the world sheet.
The explicit forms of the conjugate momenta are 
\begin{eqnarray}
{\Pi_X}_\mu&=&\frac{1}{2\pi\alpha'}\Bigl[\eta_{\mu\nu}\partial_\tau X^\nu-(A_\mu^{(0)}\hat{\delta}(\sigma-0)-A_\mu^{(\pi)}\hat{\delta}(\sigma-\pi))\Bigr],\nonumber\\
{\Pi_\psi}_\mu&=&\frac{1}{4\pi\alpha'}\bar{\psi}^\nu\gamma^0\eta_{\mu\nu}.
\end{eqnarray}

Naively thinking, one might worry about any possible $\tau$-dependency of the resulting symplectic form but the relations (\ref{rel1}) and (\ref{rel2}) and their fermion correspondents ensure its absence. Indeed one can reach to the following simple results for the bosonic part $\Omega_{\text B}$ and the fermionic part $\Omega_{\text F}$;
\begin{eqnarray}\label{symplectic}
\Omega_{\text B}&=&-\frac{1}{4\pi\alpha'}\left(B^{(0)}-B^{(\pi)}\right)_{\mu\nu}\delta x^\mu\wedge\delta x^\nu
-\frac{1}{\alpha'}\left(\eta-B^{(\pi)}\eta^{-1}B^{(0)}\right)_{\mu\nu}~~\delta x^\mu\wedge\delta a^\nu_0\nonumber\\
&&+\sum_{n\ne0}\frac{i}{n\alpha'}\left(\eta-B^{(0)}\eta^{-1}B^{(0)}\right)_{\mu\nu}\delta a^\mu_{-n}\wedge\delta a^\nu_n\nonumber\\
&&+\sum_n\frac{i}{2(n+\kappa)\alpha'}\left(\eta-B^{(0)}\eta^{-1}B^{(0)}\right)_{\mu\nu}\delta \bar{a}^\mu_{-n-\kappa}\wedge\delta a^\nu_{n+\kappa},\nonumber\\
\Omega_{\text F}&=&\sum_r\frac{i}{2\alpha'}\left(\eta-B^{(0)}\eta^{-1}B^{(0)}\right)_{\mu\nu}\left[\delta h^\mu_{-r}\wedge\delta h^\nu_r+\frac{1}{2}\delta \bar{h}^\mu_{-r-\kappa}\wedge\delta h^\nu_{r+\kappa}\right].
\end{eqnarray} 
Here, the relations (\ref{rel1}) and (\ref{rel2}) and similar relations for the fermion part should be assumed. {\em After all those constraints are worked out}, the reduced phase space variables will be composed of $\{x^0,\,x^1,\,x^2,\,a^0_n,\,a^0_{n+\kappa},\,\bar{a}^0_{-n-\kappa},\,h^0_{r},\,h^0_{r+\kappa},\,\bar{h}^0_{-r-\kappa}\}$. We recall that given the symplectic two form $\Omega=\Omega_{MN}\delta q^M\wedge \delta q^N/2$, Poisson bracket of the functions, $f(q)$ and $g(q)$, on the phase space, can be written as
$\{f,\,g\}=\Omega^{MN}\partial_Mf\,\partial_Ng$, where $\Omega^{MN}\Omega_{NL}=\delta^M_L$. Therefore by inverting the component matrix in Eq. (\ref{symplectic}), one can immediately obtain Dirac brackets, thus the (anti-)commutators in the canonical way. The results are
\begin{eqnarray}\label{commut1}
&&\left[x^0,\,x^1\right]=\frac{2\pi i\alpha'e'\sin\theta\left(1-ee'\cos\theta-bb'\right)}{\left(1-ee'\cos\theta-bb'\right)^2~\beta^2},\nonumber\\
&&\left[x^0,\,x^2\right]=\frac{-2\pi i\alpha'\left(e(1-{e'}^2+{b'}^2)-e'\cos\theta(1-ee'\cos\theta-bb')\right)}{(1-ee'\cos\theta-bb')^2~\beta^2},\nonumber\\
&&\left[x^1,\,x^2\right]=\frac{-2\pi i\alpha'\left(b(1-{e'}^2+{b'}^2)+b'(1-ee'\cos\theta-bb')\right)}{(1-ee'\cos\theta-bb')^2~\beta^2},\nonumber\\
&&\left[a^0_0,\,x^0\right]=\frac{i\alpha'{(b+b')}^2}{\left(1-ee'\cos\theta-bb'\right)^2~\beta^2},\nonumber\\
&&\left[a^0_0,\,x^1\right]=\frac{i\alpha'(b+b')(e-e'\cos\theta)}{(1-ee'\cos\theta-bb')^2~\beta^2},\\
&&\left[a^0_0,\,x^2\right]=\frac{i\alpha'(b+b')e'\sin\theta}{(1-ee'\cos\theta-bb')^2~\beta^2},\nonumber\\
&&\left[a^0_m,\,a^0_n\right]=\frac{-\alpha'{(b+b')}^2~~m~\delta_{n+m}}{2{(1-ee'\cos\theta-bb')}^2~\beta^2},\nonumber\\
&&\left[\bar{a}^0_{m-\kappa},\,a^0_{n+\kappa}\right]=-\frac{2\alpha'(p^2+q^2)}{\Omega''}~~(m-\kappa)~\delta_{m+n},\nonumber\\
&&\left\{h^0_r,\,h^0_s\right\}=-\frac{\alpha'{(b+b')}^2}{{(1-ee'\cos\theta-bb')}^2~\beta^2}\delta_{r+s},\nonumber\\
&&\left\{\bar{h}^0_{s-\kappa},\,h^0_{r+\kappa}\right\}=-\frac{4\alpha'(p^2+q^2)}{\Omega''}\delta_{r+s}
\nonumber
\end{eqnarray}
where $\Omega''\equiv -e^2(p^2+q^2-t^2-u^2)-(b^2+1)(r^2+s^2+t^2+u^2)+p^2+q^2+2eb(rp+sq)$. Although it is far from clear whether $\Omega''$ is positive for general value of $\beta$, one can see it becomes $-2(1+b^2)(1-e^2)$, thus negative in the special case of $e'=e, \, b'=b, \, \theta=0$. (This special case was considered in Ref. \cite{bak}.) From hereon, we assume that it is negative, otherwise, the Hamiltonian becomes negative definite and unbounded below. This case will be dealt in a forthcoming paper \cite{future}.

The commutators can be written in the canonical fashion by the following normalizations;
\begin{eqnarray}
\alpha^0_{\pm n}&=&\sqrt{2{(1-ee'\cos\theta-bb')}^2~\beta^2\over\alpha'{(b+b')}^2}~a^0_{\pm n},\nonumber\\
\alpha^0_{\pm n\pm\kappa}\,\,(\bar\alpha^0_{\pm n\pm\kappa})&=&\sqrt{-\Omega''\over2\alpha'(p^2+q^2)}~a^0_{\pm n\pm\kappa}\,\,(\bar a^0_{\pm n\pm\kappa}),\nonumber\\
\varphi^0_{\pm r}&=&\sqrt{{(1-ee'\cos\theta-bb')}^2~\beta^2\over\alpha'{(b+b')}^2}~h^0_{\pm r},\nonumber\\
\varphi^0_{\pm r\pm\kappa}\,\,(\bar\varphi^0_{\pm r\pm\kappa})&=&\sqrt{-\Omega''\over4\alpha'(p^2+q^2)}~h^0_{\pm r\pm\kappa}\,\,(\bar h^0_{\pm r\pm\kappa}),
\end{eqnarray}

by which
\begin{eqnarray}
\Bigl[\alpha^0_m,~~\alpha^0_n\Bigr]&=&-m\delta_{m+n},\qquad
\Bigl[\bar{\alpha}^0_{m-\kappa},~~\alpha^0_{n+\kappa}\Bigr]=(m-\kappa)\delta_{m+n},\nonumber\\
\Bigl\{\varphi^0_r,~~\varphi^0_s\Bigr\}&=&-\delta_{r+s},\qquad\quad~
\Bigl\{\bar{\varphi}^0_{r-\kappa},~~\varphi^0_{s+\kappa}\Bigr\}=\delta_{r+s}.
\end{eqnarray}
(From hereon, we omit the recurrent superscripts $0$'s.)
 
\section{Noncommutative Space and Time}

The commutators among $X^{0,1,2}$ vanish in the bulk, i.e, $0<\sigma,\,\sigma'<\pi$. To illustrate this, let us consider $[X^1(\sigma),\,X^2(\sigma')]$ for example. After a lengthy calculation, all the coordinate dependent terms are cancelled off and the following terms remain;
\begin{eqnarray}
&&[X^1(\sigma),\,X^2(\sigma')]\\
&&\quad=\left\{
\begin{array}{ll}
\vspace{0.8cm}
[x^1,\,x^2]-2\pi i\alpha'\left[
\frac{be'^2\sin^2{\theta}}{\left(1-ee'\cos{\theta-bb'}\right)^2\beta^2}+2\frac{st-ru+b\left(t^2+u^2\right)\beta}{\Omega''\beta}\right]&(0<\sigma-\sigma'\leq\pi)\cr
[x^1,\,x^2]-2\pi i\alpha'\left[-\frac{\left(e-e'\cos{\theta}\right)\left(b'e+be'\cos{\theta}\right)}{\left(1-ee'\cos{\theta-bb'}\right)^2\beta^2}+2\frac{st-ru+\left(b\left(r^2+s^2\right)-e\left(rp+sq\right)\right)\beta}{\Omega''\beta}\right]&(-\pi\leq\sigma-\sigma'<0).\nonumber
\end{array}
\right.
\end{eqnarray} 
The second and the third terms came from the commutators $[a_n,\,a_m]$ and $[a_{n+\kappa},\,\bar{a}_{m-\kappa}]$ respectively. The above commutator vanishes upon the imposition of the result of $[x^1,\,x^2]$ obtained in (\ref{commut1}). This is consistent with our conception that the gauge fields coupling to the string end points do not affect the string interior. Therefore as far as these 
 commutators $[x,\,x]$'s are concerned, one can determine them by requiring the vanishing of the commutators $[X,\,X]$ in the bulk, as was done in Ref. \cite{bak}.

Since the string end points are coupled to both electric and magnetic fields, we expect the space-time and the space-space noncommutativities at the worldsheet boundaries. We omit the cumbersome computational details but would like to mention one technical point of what causes the discrepancy between the bulk commutators and the boundary commutators. The typical expressions we would encounter in the calculation of the commutators, are of the form
\begin{eqnarray}
\sum\limits_{n=-\infty}^{\infty}\frac{\sin{n\vartheta}}{n+\kappa}=\left\{\begin{array}{ll} \frac{\pi}{\sin{\kappa\pi}}\sin{\kappa(\pi-\vartheta)}&\quad(0<\vartheta<2\pi)\cr
0&\quad(\vartheta=0)\end{array}\right.,
\end{eqnarray}
which singles out the point $\vartheta=\sigma-\sigma'=0$.
Let us sum up the result:
\begin{eqnarray}
&&\left.\Bigl[X^0(\sigma),X^1(\sigma')\Bigr]\right|_{\sigma=\sigma'=0}=0\nonumber\\
&&\left.\Bigl[X^0(\sigma),X^1(\sigma')\Bigr]\right|_{\sigma=\sigma'=\pi}=2\pi i\alpha'\frac{e'\sin\theta}{1-e'^2+b'^2}\nonumber\\
&&\left.\Bigl[X^0(\sigma),X^2(\sigma')\Bigr]\right|_{\sigma=\sigma'=0}=-2\pi i\alpha'\frac{e}{1-e^2+b^2}\nonumber\\
&&\left.\Bigl[X^0(\sigma),X^2(\sigma')\Bigr]\right|_{\sigma=\sigma'=\pi}=2\pi i\alpha'\frac{e'\cos\theta}{1-e'^2+b'^2}\\
&&\left.\Bigl[X^1(\sigma),X^2(\sigma')\Bigr]\right|_{\sigma=\sigma'=0}=-2\pi i\alpha'\frac{b}{1-e^2+b^2}\nonumber\\
&&\left.\Bigl[X^1(\sigma),X^2(\sigma')\Bigr]\right|_{\sigma=\sigma'=\pi}=-2\pi i\alpha'\frac{b'}{1-e'^2+b'^2}.\nonumber
\end{eqnarray}
All of them can be recast into the following covariant form
\begin{eqnarray}
\left.\Bigl[X^\mu(\sigma),X^\nu(\sigma')\Bigr]\right|_{\sigma=\sigma'=(1\mp1)\pi/2}=\left.\pm4\pi i\alpha'\frac{B^{\mu\nu}}{2+B_{\lambda\rho}B^{\lambda\rho}}\right|_{\sigma=\sigma'=(1\mp1)\pi/2}.
\end{eqnarray}

\section{How to Evade Tachyons}

In this section, we explore the condition for the tachyons to disappear from the string spectrum. Let us first look at the Hamiltonian.
\begin{eqnarray}
H&=&H_{\text B}+H_{\text F}\nonumber\\
&=&\frac{1}{2\pi\alpha'}\int\limits^\pi_0 d\sigma\left(\partial_+ X^\mu\partial_+ X^\nu+\partial_- X^\mu\partial_- X^\nu\right)\eta_{\mu\nu}-\frac{i}{4\pi\alpha'}\int\limits^\pi_0 d\sigma\left(\psi^\mu_-\partial_\sigma\psi^\nu_--\psi^\mu_+\partial_\sigma\psi^\nu_+\right)\nonumber\\
&=&-\frac{1}{2}\sum\limits_{n}\alpha_{-n}\alpha_{n}+\frac{1}{2}\sum\limits_n\left(\alpha_{-n+\kappa}\bar{\alpha}_{n-\kappa}+\bar{\alpha}_{-n-\kappa}\alpha_{n+\kappa}\right)\nonumber\\
&&-\frac{1}{2}\sum_rr\varphi_{r}\varphi_{-r}+\frac{1}{2}\sum_r\left(r+\kappa\right)\left[-\varphi_{r+\kappa}\bar\varphi_{-r-\kappa}+\bar\varphi_{-r-\kappa}\varphi_{r+\kappa}\right].
\end{eqnarray}
Here we observe a novel transfer of degrees of freedom. We started from a string living in $(2+1)$-dimensions. (Let aside the other transverse seven dimensions for the time being.) In order for the interstring to satisfy the generic boundary conditions at both ends, the modes in three directions ought to be related by the relations (\ref{rel1}) and (\ref{rel2}). We used up those relations and worked with the modes in the time direction. Although the integer modes of thus remained temporal component make ghost states (which are to be offset by the states coming from one of the transverse seven directions), the peculiar fractional modes fill up the vacancy of the spatial directions. 

\subsection{the Zero Point Energy}

The fractional modes affect the value of zero point energy in the NS-sector.
With the negative modes as the raising operators (and the opposite for the integer modes), one can recast the above Hamiltonian in the normal ordered fashion,
\begin{eqnarray}
H&=&-\frac{1}{2}\alpha_0{}^2-\sum\limits_{n>0}\alpha_{n}\alpha_{-n}+\sum\limits_{n>0}\alpha_{-n+\kappa}\bar{\alpha}_{n-\kappa}+\sum\limits_{n\ge0}\bar{\alpha}_{-n-\kappa}\alpha_{n+\kappa}\nonumber\\
&&-\sum\limits_{r>0}r\varphi_{r}\varphi_{-r}+\sum\limits_{r>0}(r-\kappa)\varphi_{-r+\kappa}\bar\varphi_{r-\kappa}+\sum\limits_{r\ge0}(r+\kappa)\bar\varphi_{-r-\kappa}\varphi_{r+\kappa}\\
&&-\frac{1}{2}\sum\limits_{n>0}n+\frac{1}{2}\sum\limits_{n>0}\left(n-\kappa\right)+\frac{1}{2}\sum\limits_{n\ge0}\left(n+\kappa\right)+\frac{1}{2}\sum\limits_{r>0}r-\frac{1}{2}\sum\limits_{r>0}\left(r-\kappa\right)-\frac{1}{2}\sum\limits_{r\ge0}\left(r+\kappa\right).\nonumber
\end{eqnarray}
The last line collects the zero point energies for $(2+1)$-dimensional part over which nontrivial gauge fields are laid. The full zero point energy of the interstring living in $(9+1)$-dimensions will be
\begin{eqnarray}
E_0&=&\frac{1}{2}\left(-\sum\limits_{n>0}n+\sum\limits_{n>0}\left(n-\kappa\right)+\sum\limits_{n>0}\left(n-1+\kappa\right)+7\sum\limits_{n>0}n\right)\nonumber\\
&&-\frac{1}{2}\left(-\sum\limits_{r>0}r+\sum\limits_{r>0}\left(r-\kappa\right)+\sum\limits_{r>0}\left(r-1+\kappa\right)+7\sum\limits_{r>0}r\right)\nonumber\\
&=&\left\{
\begin{array}{lr}0&\quad({\text R})\\
 -\frac{1}{4}-\frac{\vert\kappa-\frac{1}{2}\vert}{2}&\quad({\text NS}) 
\end{array}\right.
\end{eqnarray}

It is important to see that the point $\kappa=1/2$ does not signify the seeming role exchange between the NS-sector and the R-sector. This is because the parameter $\kappa$ affects not only the fermionic part but also the bosonic part.

\subsection{NS-spectrum}

The NS-sector is sensitive to the background gauge fields and its ground state has negative energy $E_0$ at the generic value of $\kappa$. Over this ground state, the first pair of the creation operators $(\varphi,\,\bar{\varphi})$ will make the next two states with energies
\begin{eqnarray}
E_1=E_0+\left|\kappa-\frac{1}{2}\right|,\qquad E_2=E_0-\left|\kappa-\frac{1}{2}\right|+1.
\end{eqnarray}
Fig. 2 shows the spectrum of lower energy states in the NS-sector. All the energy levels reveal a sharp cusp at $\kappa=1/2$, at which $\beta=\tan{\kappa\pi}$ changes its sign. Let us denote the NS-ground state at $\kappa$ as $|0>_\kappa$. The first and second excited states are obtained by acting $\varphi_{-\frac{1}{2}+\kappa}$ and $\bar{\varphi}_{-\frac{1}{2}-\kappa}$ respectively on $|0>_\kappa$ when $0<\kappa<1/2$, while $\bar{\varphi}_{\frac{1}{2}-\kappa}$, $\varphi_{-\frac{3}{2}+\kappa}$ when $1/2<\kappa<1$. 

\begin{figure}\label{fig2}
\epsfbox{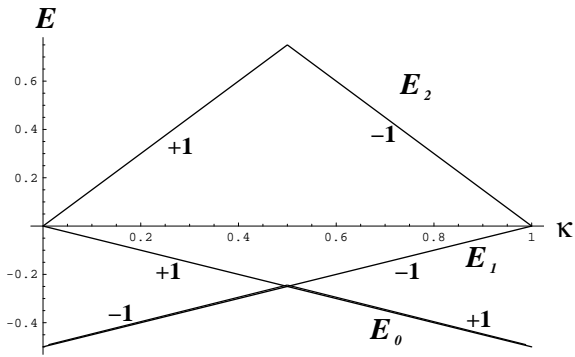}
\caption{The plot shows the energy spectrum in the NS-sector. It varies with the mode number $\kappa$. The numbers $\pm1$ denote G-parity charges, $(-)^F$, of the states. As for the D2-D2 pair, GSO projection keeps only the states with the even G-parity, ($G=+1$), while for the D2-\d2bar pair, the states with the odd G-parity, ($G=-1$), will be kept.}
\end{figure}

We observe that the energies of the first and the second excited states merge to be equal to zero at two special points, $\kappa=0$ and $1$. Removing the unwanted tachyonic NS ground state $|0>_\kappa$ via the GSO prescription, we would get tachyon-free spectrum at these two points. The massless states compose ${\bf 2_v}$ of $SO(2)$. The rest transverse components of creation operators, $\psi^i_{-1/2}\,(i=3,\cdots,8)$ will make ${\bf 6_v}$ of $SO(6)$. Away from these two points, $\kappa=0$ and $1$, the tachyonic states would survive GSO projection and trigger the instability of the system.

In view of the definition of $\beta=\tan{\kappa\pi}$ and the relation of (\ref{kappa}), we note that there could be many different sets of the electric fields and the magnetic fields corresponding to those two special points. For example, both the following two sets of background fields, $\vec{e}=\vec{e}'=0$, $\vec{b}=\vec{b}'=0$, and $\vec{e}=\vec{e}'=\vec{1}$, $\vec{b}+\vec{b}'=0$ give the same value of $\beta=0$, therefore correspond to the points, $\kappa=0$ or $1$. In fact, the D2-D2 pair in the trivial background fields and the D2-\d2bar pair in the latter set of the background fields are two typical supersymmetric configurations (preserving $16$ and $8$ supersymmetries respectively), thus their interstrings will be tachyon-free. The drastic sign change in $\beta$ and the sharp cusp of all the energy levels at $\kappa=1/2$ suggest that the two configurations correspond to different points of the values, $\kappa=0$ and $1$. Here, one could ask `which configuration to which point?'

The two points $\kappa=0, 1$ are actually equivalent in the sense that the Hamiltonian and the commutators among oscillators are invariant under the following $Z\!\!\!\!Z_{2}$ mapping:
\begin{eqnarray}\label{zz}
&&\kappa\Longrightarrow1-\kappa',\qquad r\Longrightarrow-(1+r'),\nonumber \\
&&\varphi_{r+\kappa}\Longrightarrow\bar{\varphi}'_{-r'-\kappa'}\,,\quad
\bar{\varphi}_{-r-\kappa}\Longrightarrow\varphi'_{r'+\kappa'}\,.  
\end{eqnarray}
This implies the details of the correspondence between the aforementioned two supersymmetric configurations and the points $\kappa=0, 1$ is just a matter of convention. Henceforth, we consider the trivial background is pertaining to the point $\kappa=0$ without loss of generality. 

The notion of the D2-D2 pair and the D2-\d2bar pair is only determined by the way of the GSO prescription. In the standard rule, the states with the odd fermion number, $F$, are projected out for the D2-D2 pair, while the opposite way is applied for the D2-\d2bar pair. According to our convention in the above, we assign odd/even fermion number to the NS ground state in the region $(0<\kappa<1/2)$/($1/2<\kappa<1$). As for the D2-D2 pair, only the states with the even G-parity ($G=(-1)^{F}=1$) will survive upon the GSO projection. As a consequence, the D2-D2 pair becomes free from the tachyons in the trivial background ($\kappa=0$). As for the D2-\d2bar pair, the GSO projection keeps only the states with the odd G-parity, so that the tachyonic modes go away when $\vec{e}=\vec{e}'=\vec{1}$, $\vec{b}+\vec{b}'=0$, that is at $\kappa=1$.

\subsection{R-spectrum}

Let us turn to the R-sector. The zero modes $\varphi_0,\,\psi^i_0, \,(i=3,\cdots,9)$, generate eight ($2^{8/2-1}$) massless {\em physical} states, which are massless fermions in the spinor representations, ${\bf 4}+{\bf 4'}$ of the transverse symmetry, $SO(6)\simeq SU(4)$. Especially in $\kappa\rightarrow0$ limit, the operators, $\varphi_\kappa, \,\bar{\varphi}_{-\kappa}$ will make an extra 2-fold degeneracy corresponding to ${\bf 1}+{\bf 1'}$ of $SO(2)\simeq U(1)$. Therefore on the whole, in $\kappa\rightarrow0$ limit, 16 massless physical states will be made in the R-sector, which are in $\left({\bf 1}+{\bf 1'}\right)\otimes\left({\bf 4}+{\bf 4'}\right)$. The GSO prescription for the R-sector keeps either one part of the spinors ${\bf 1}\otimes{\bf 4}+{\bf 1'}\otimes{\bf 4'}$ or ${\bf 1}\otimes{\bf 4'}+{\bf 1'}\otimes{\bf 4}$. Hence, thus remained eight massless fermionic degrees match exactly with those of eight bosons in ${\bf 2_v}+{\bf 6_v}$. This opens up the possibility of any preserved supersymmetry. (It should be noted that the match of the bosonic and the fermionic degrees just at the massless level is not sufficient for the supersymmetry.) In the limit $\kappa\rightarrow1$, one can reach the same conclusion by using the $Z\!\!\!\!Z_{2}$ mapping of Eq. (\ref{zz}). Consequently, in the limit of $\kappa=0$ or $1$, thus Eq. (\ref{notachyon}), the bosonic and the fermionic degrees of freedom match exactly at least at the massless level.

The precise number of supersymmetries preserved at $\kappa=0,\,1$ is hard to tell in this context. The tachyon removal condition, (\ref{notachyon}) is only the necessary condition for the supersymmetry. For example when $\vec{e}'=\vec{e}$ and $\vec{b}'=\lambda\vec{b}$, the condition says $\left(1-e^2\right) \left(\lambda-1\right)^2 b^2=0$. We know of the case $e=1$ and $\lambda=1$ preserving 16 supersymmetries. However, the case where $e=1$ and $\lambda\ne1$ preserves only 8 supersymmetries. See Ref. \cite{cho-oh} for details. 

\section{Lineal Motion of the Interstrings}

Since the zero modes $\{a^0_0,\, a^1_0,\, a^0_2\}$ are not independent, it is tempting to see the motion of the interstring. Let us calculate the momentum of its center of mass. Under the gauge $A_{\mu}=-B_{\mu\nu}X^\nu$, the integration of ${\Pi_X}_\mu$ results in
\begin{eqnarray}
P_\mu&=&\frac{1}{2\pi\alpha'}\int\limits^\pi_0d\sigma\,\,\partial_\tau X_\mu  -\frac{1}{2\pi\alpha'}\left(A^{(0)}_\mu-A^{(\pi)}_\mu\right)\nonumber\\
&=&\frac{1}{2\pi\alpha'}\left[2\pi\eta_{\mu\nu}a^\nu_0-2\pi B^{(\pi)}_{\mu\rho}\eta^{\rho\lambda}B^{(0)}_{\lambda\nu}a^\nu_0+ \left(B^{(0)}_{\mu\nu}-B^{(\pi)}_{\mu\nu}\right)x^\nu\right].
\end{eqnarray}
The first term in the second line, coming from $\partial_\tau X_\mu$, corresponds to the kinetic momentum $k_\mu$, which differs from the conjugate momentum $P_\mu$ in the canonical way in the presence of gauge fields. 
The relation (\ref{rel1}) among the zero modes determines the kinetic momentum as
\begin{eqnarray}\label{momentum}
\left(k^\mu\right)=\frac{a_0}{\alpha'}\left(1,\,\frac{e-e'\cos{\theta}}{b+b'},\,\frac{e'\sin{\theta}}{b+b'}\right).
\end{eqnarray}
Therefore the interstring moves in the direction normal (in the flat metric) to the vector $\vec{e}-\vec{e}'$ as is shown in Fig. 3. To be more precise, the motions are classified into two cases:
i) $\vec{e}=\vec{e}',\quad \vec{b}\ne\vec{b}'$ : static,
ii) $\vec{e}\ne\vec{e}',\quad \vec{b}\ne\vec{b}'$ : lineal motion. 

\begin{figure}\label{fig2}
\epsfbox{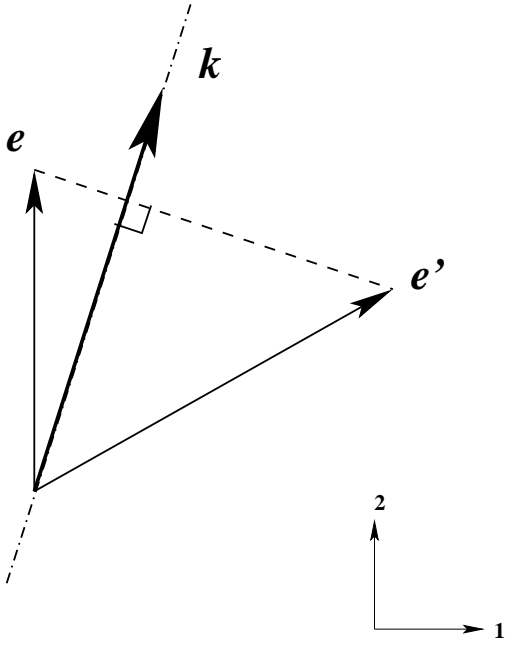}
\caption{The interstring over a D2-\d2bar pair can move only along the direction normal to the vector $\vec{e}-\vec{e}'$.}
\end{figure}

So far we have considered a D-$\overline{\text D}$ string that runs from the D2-brane to the \d2bar-brane. All the D-$\overline{\text D}$ strings will move in parallel with the same momentum given in Eq. (\ref{momentum}). The same is true for the $\overline{\text D}$-D strings: they follow the same parallel trails with the same momentum as those of D-$\overline{\text D}$ strings. This latter fact can be seen easily by looking at the action, (\ref{action}). A $\overline{\text D}$-D string will be described by the same action but with the exchange of $A^{(0)}_\mu$ for $A^{(\pi)}_\mu$ and vice versa. In (\ref{rel1}), this amounts to the overall sign flip, which will not make any change in the above results. 

Some comments on the magnitude of momentum are in order. 
The magnitude of the momentum depends crucially on whether we use the `open string metric' $G\equiv E^{T}\eta^{-1} E$ (with the relations, (\ref{rel1}) and (\ref{rel2}) assumed) or the `closed string metric' $\eta$. 
In the open string metric, it is
\begin{eqnarray}
k^\mu G_{\mu\nu}k^\nu=-\frac{\beta^2\left(1-\vec{e}\cdot\vec{e}'+\vec{b}\cdot\vec{b}'\right)^2}{\alpha'^2\left(\vec{b}-\vec{b}'\right)^2}\left(a_0\right)^2=-\frac{1}{2\alpha'}\left(\alpha_0\right)^2.
\end{eqnarray}
Since $\left(\alpha_0\right)^2=2E_0$ for physical state satisfying Virasoro constraints, the center of mass momentum becomes spacelike when the interstring is in its NS ground state $|0;k>_{\kappa}$. This is what we usually understand about the tachyons moving faster than the speed of light. 

Meanwhile in the closed string metric, the momentum has the magnitude,
\begin{eqnarray}
k^\mu\eta_{\mu\nu}k^\nu=\frac{\left(\vec{e}-\vec{e}'\right)^2-\left(\vec{b}-\vec{b}'\right)^2}{\alpha'^2\left(\vec{b}-\vec{b}'\right)^2}\left(a_0\right)^2=\frac{\left(\vec{e}-\vec{e}'\right)^2-\left(\vec{b}-\vec{b}'\right)^2}{2\alpha'\beta^2\left(1-\vec{e}\cdot\vec{e}'+\vec{b}\cdot\vec{b}'\right)^2}\left(\alpha_0\right)^2.
\end{eqnarray}
It is very amusing to see whether, in some regime of gauge fields, the value of $(\vec{b}-\vec{b}')^2$ can be less than that of $\left(\vec{e}-\vec{e}'\right)^2$. If it is, the momentum squared in the open string metric and in the closed string metric have different signatures in such a region. In a sense, this look like a feature that is salient to the black holes; the asymptotic timelike Killing vector becomes spacelike inside the event horizon. However, it is not likely to be so. It is improbable to satisfy the above condition and the assumptions, $\beta^2>0$ and $\Omega''<0$ at the same time. Actually we could not find any single example of this case.
  
\section{Discussions}

Let us specialize our result of the tachyon removal condition, $\beta^2\rightarrow0$ to some known cases of supersymmetric configurations of D-branes. As for the D2-\d2bar pair considered in Ref. \cite{bak1} and Ref. \cite{bak}, $e=e'=1$, $b=b'$, and $\theta=0$. Therefore $\beta^2=0$. Another interesting example is the null scissors configuration discussed in Ref. \cite{bachas}. It is a system composed of one static tilted D-string and another vertical D-string in relative motion with respect to the first one. (See Fig. 4.) 

\begin{figure}\label{fig4}
\epsfbox{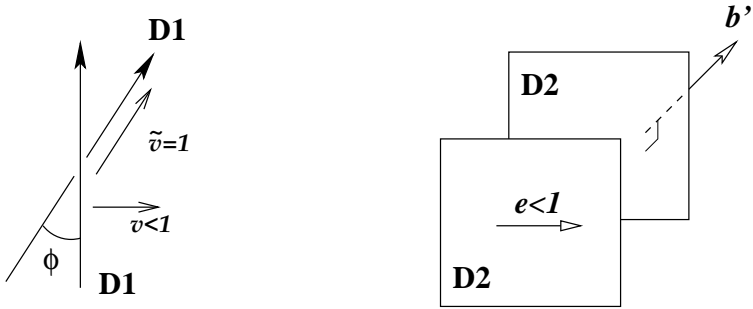}
\caption{Null scissors preserve 8 supersymmetries. As the vertical D-strings moves right, the intersecting point moves along the tilted D-string with the speed of light. In the T-dual picture, the system is composed of two D2-branes one of which has the noncritical electric field only and the other has the magnetic field only.}
\end{figure}

The system preserve eight supersymmetries when the intersecting point moves at  the speed of light. The same physics can be read off in the T-dualized configuration, which is nothing but a D2-D2 pair with $e=v,\,e'=0$ and $b=0,\,b'=\tan{\phi}$. See \cite{cho-oh} for details. The tachyon removal condition reads as
\begin{eqnarray}
\beta^2=\left(1-e^2\right)\left(1+b^2\right)-1^2=-v^2\sec^2{\phi}+\tan^2{\phi}=\frac{1}{\cos^2{\phi}}\left(-v^2+\sin^2{\phi}\right)\rightarrow0,
\end{eqnarray}
which corresponds to the condition for the intersecting point of the scissors to move with the speed of light.

We conclude this paper with some remarks on future works to be done regarding this work. We have assumed that $\beta^2>0$ or $\Omega''<0$. Actually the specific case where $e'=e,\,b'=b,$ and $\theta=0$ satisfies those conditions. In the region of the background fields where some of these conditions are not valid, we would meet some ghost excitations due to the wrong signature of the commutators. It would be interesting to explorer the physics in such a case \cite{future}. It would be also very interesting to see the one loop interaction of the D2- and the \d2bar-brane. According to our result, all the interstrings make lineal trails on the (anti-)D-brane world volume. The splitting or joining of strings is highly suppressed on shell. This tree level behavior might constrain the one loop motion. 

\bigskip

\acknowledgments

We thank Dongsu Bak and Nobuyoshi Ohta for useful discussions. JHC is supported in part by KRF through Project No. 2003-070-C00011.

\end{document}